\documentclass[prd,twocolumn,groupedaddress,showpacs,nofootinbib,floatfix]{revtex4}
\usepackage{graphicx}
\usepackage{dcolumn}
\usepackage{amssymb}
\usepackage{mathrsfs}
\usepackage{amsmath}
\usepackage{epsfig}
\usepackage[dvips]{color}
\usepackage{hhline}

\begin{document}

\title{A new leptogenesis scenario parametrized by Dirac neutrino mass matrix}

\author{Pei-Hong Gu}
\email{peihong.gu@sjtu.edu.cn}

\affiliation{Department of Physics and Astronomy, Shanghai Jiao Tong
University, 800 Dongchuan Road, Shanghai 200240, China}

\begin{abstract}

In an $SU(3)^{}_{c}\times SU(2)^{}_{L}\times SU(2)^{}_{R}\times U(1)^{}_{B-L}$ left-right symmetric framework, we present a new leptogenesis scenario parametrized by Dirac neutrino mass matrix. Benefited from the parity symmetry motivated to solve the strong CP problem, the dimensionless couplings of the mirror fields are identified to those of the ordinary fields. In particular, the mirror Dirac neutrinos have a heavy mass matrix proportional to the light mass matrix of the ordinary Dirac neutrinos. Through the $SU(2)_R^{}$ gauge interactions, the mirror neutrinos can decay to generate a lepton asymmetry in the mirror muons and an opposite lepton asymmetry in the mirror electrons. Before the $SU(2)_L^{}$ sphaleron processes stop working, the mirror muons can efficiently decay into the ordinary right-handed leptons with a dark matter scalar and hence the mirror muon asymmetry can be partially converted to a desired baryon asymmetry.

\end{abstract}


\maketitle

\section{Introduction}

People have proposed various baryogensis mechanisms to understand the cosmic matter-antimatter asymmetry which is the same as a baryon asymmetry. The leptogenesis \cite{fy1986,lpy1986,fps1995,ms1998,bcst1999,hambye2001,di2002,gnrrs2003,hs2004,bbp2005,dnn2008} in seesaw \cite{minkowski1977,mw1980} context has become one of the most attractive baryogenesis mechanisms because it can simultaneously explain the generation of baryon asymmetry and the smallness of neutrino masses. However, the conventional leptogenesis scenario contains many free parameters so that it cannot give a distinct relation between the baryon asymmetry and the neutrino mass matrix unless we do some assumptions on the texture of the relevant masses and couplings. For example, we can expect a successful leptogenesis in the canonical seesaw model even if the neutrino mass matrix does not contain any CP phases \cite{di2001}.

In this paper we shall propose a new leptogenesis scenario parametrized by Dirac neutrino mass matrix in an $SU(3)^{}_{c}\times SU(2)^{}_{L}\times SU(2)^{}_{R}\times U(1)^{}_{B-L}$ left-right symmetric \cite{ps1974} model.
Motivated by solving the strong CP problem without introducing an unobserved axion \cite{ms1978,bm1989,bcs1991,lavoura1997,gu2012,cgnr2013}, we will consider a parity symmetry under which the dimensionless couplings of the mirror fields are identified to those of the ordinary fields. Through the $SU(2)_R^{}$ gauge interactions, the mirror Dirac neutrinos, which have a heavy mass matrix proportional to the seesaw-suppressed \cite{gh2006} mass matrix of the ordinary Dirac neutrinos, can decay to produce a lepton asymmetry in the mirror muons and an opposite lepton asymmetry in the mirror electrons. Both of the mirror muons and electrons can decay into the ordinary right-handed leptons with a dark matter scalar. However, the lifetime of the mirror muons can be much shorter than that of the mirror electrons. This means the mirror electron asymmetry can be expected not to participate in the $SU(2)_L^{}$ sphaleron processes. Instead, only the mirror muon asymmetry can be partially converted to a baryon asymmetry.

\section{The model}

\begin{table}
\vspace{0.25cm}
\begin{center}
\begin{tabular}{|c|c|c|c|c|}  \hline &&&&\\[-2.5mm]
$~~\textrm{Field}~~$ &~~$Z_4^{}$~~ & ~~$Z_3^{}$~~& $~~Z'^{}_4~~$ & $~~Z'^{}_{3}~~$\\
&&&&\\[-3.0mm]
\hline \hline&&&& \\[-1.5mm]$q_L^{}(3,2,1,+\frac{1}{6})$ &$+1$  & $e^{i2\pi/3}_{}$&&\\
&&&& \\[-1.5mm]
\hline&&&&\\[-1.5mm]$d_R^{}(3,1,1,-\frac{1}{3})$ & $+i$  &$e^{i2\pi/3}_{}$&&\\
&&&& \\[-1.5mm]
\hline&&&& \\[-1.5mm]$u_R^{}(3,1,1,+\frac{2}{3})$  & $+i$& $e^{i2\pi/3}_{}$ &&\\
&&&&\\[-1.5mm]
\hline&&&& \\[-1.5mm]$l_L^{}(1,2,1,-\frac{1}{2})$ & $+1$  &$e^{i4\pi/3}_{}$ &&\\
&&&& \\[-1.5mm]
\hline&&&& \\[-1.5mm]$e_R^{}(1,1,1,-1)$ & $-1$  &$e^{i4\pi/3}_{}$ &&\\
&&&&\\[-1.5mm]
\hline&&&& \\[-1.5mm]$\nu_R^{}(1,1,1,~~0)$ & $+1$  &$e^{i4\pi/3}_{}$ &&\\
&&&&\\[-1.5mm]
\hline&&&& \\[-1.5mm]$\phi_d^{}(1,2,1,-\frac{1}{2})$ & $+i$  &$1$ &&\\
&&&&\\[-1.5mm]
\hline&&&& \\[-1.5mm]$\phi_u^{}(1,2,1,-\frac{1}{2})$ & $-i$  &$1$ &&\\
&&&&\\[-1.5mm]
\hline&&&& \\[-1.5mm]$\phi_e^{}(1,2,1,-\frac{1}{2})$ & $-1$  &$1$ &&\\
&&&&\\[-1.5mm]
\hline&&&& \\[-1.5mm]$\phi_\nu^{}(1,2,1,-\frac{1}{2})$ & $+1$  &$1$ &&\\
&&&&\\[-1.5mm]
\hline \hline&&&& \\[-1.5mm]$q'^{}_R(3,1,2,+\frac{1}{6})$&& &$+1$  & $e^{i4\pi/3}_{}$\\
&&&& \\[-1.5mm]
\hline&&&&\\[-1.5mm]$d'^{}_L(3,1,1,-\frac{1}{3})$ &&& $+i$  &$e^{i4\pi/3}_{}$\\
&&&& \\[-1.5mm]
\hline&&&& \\[-1.5mm]$u'^{}_L(3,1,1,+\frac{2}{3})$  && & $+i$& $e^{i4\pi/3}_{}$\\
&&&&\\[-1.5mm]
\hline&&&& \\[-1.5mm]$l'^{}_R(1,1,2,-\frac{1}{2})$ & && $+1$  &$e^{i2\pi/3}_{}$\\
&&&& \\[-1.5mm]
\hline&&&& \\[-1.5mm]$e'^{}_L(1,1,1,-1)$ &&& $-1$  &$e^{i2\pi/3}_{}$ \\
&&&&\\[-1.5mm]
\hline&&&& \\[-1.5mm]$\nu'^{}_L(1,1,1,~~0)$ &&& $+1$  &$e^{i2\pi/3}_{}$ \\
&&&&\\[-1.5mm]
\hline&&&& \\[-1.5mm]$\phi'^{}_d(1,1,2,-\frac{1}{2})$ & && $+i$  &$1$\\
&&&&\\[-1.5mm]
\hline&&&& \\[-1.5mm]$\phi'^{}_u(1,1,2,-\frac{1}{2})$ & && $-i$  &$1$\\
&&&&\\[-1.5mm]
\hline&&&& \\[-1.5mm]$\phi'^{}_e(1,1,2,-\frac{1}{2})$ & && $-1$  &$1$\\
&&&&\\[-1.5mm]
\hline&&&& \\[-1.5mm]$\phi'^{}_\nu(1,1,2,-\frac{1}{2})$ & && $+1$  &$1$\\
&&&&\\[-1.5mm]
\hline\hline&&&& \\[-1.5mm]$\chi(1,1,1,0)$ & $+i$ & $e^{i2\pi/3}_{}$  & $-i$ & $e^{i2\pi/3}_{}$\\
&&&& \\[-1.5mm]
\hline&&&&\\[-1.5mm]$\xi(1,1,1,0)$ & $-1$  &$e^{i4\pi/3}_{}$& $-1$ & $e^{i4\pi/3}_{}$\\[-1.5mm]
&&&& \\
\hline
\end{tabular}
\vspace{0.25cm}
\caption{\label{field} Fermions and scalars in the model. Here the brackets following the fields describe the transformations under the $SU(3)_c^{}\times SU(2)_L^{}\times SU(2)_R^{}\times U(1)_{B-L}^{}$ gauge groups. The ordinary(mirror) fermions and Higgs scalars without(with) prime take the charges under the $Z_4^{}\times Z_3^{}$($Z'^{}_4\times Z'^{}_3$) discrete symmetries. As for the other two scalars $\xi$ and $\chi$, they cross the ordinary and mirror sectors.}
\end{center}
\end{table}

The fermions and scalars in the model are summarized in Table \ref{field}, where the brackets following the fields describe the transformations under the $SU(3)_c^{}\times SU(2)_L^{}\times SU(2)_R^{}\times U(1)_{B-L}^{}$ gauge groups. In addition, the ordinary fermions and Higgs scalars without prime respect a $Z_4^{}\times Z_3^{}$ discrete symmetry. Accordingly, the mirror fermions and Higgs scalars with prime respect a $Z'^{}_4\times Z'^{}_3$ discrete symmetry. As for the other two scalars $\xi$ and $\chi$, they are non-trivial under both of the $Z_4^{}\times Z_3^{}$ and $Z'^{}_4\times Z'_3$ symmetries and cross the ordinary and mirror sectors.

The $Z^{}_4$ and $Z'^{}_4$ discrete symmetries are allowed to be softly broken. Therefore, the $[SU(2)^{}_R]$-doublet Higgs scalars $\phi'^{}_{d,u,e,\nu}$ can mix with each other,
\begin{eqnarray}
V\supset \sum_{a=d,u,e,\nu}^{}\mu'^2_a \phi'^\dagger_a\phi'^{}_a + \sum_{a\neq b}^{}\mu'^2_{ab} \phi'^\dagger_a \phi'^{}_b~~(\mu'^2_{ab}=\mu'^2_{ba})\,.
\end{eqnarray}
One of their linear combinations is
\begin{eqnarray}
\varphi'&=&\frac{\langle\phi'^{}_{d}\rangle \phi'^{}_{d}+\langle\phi'^{}_{u}\rangle\phi'^{}_{u}+\langle\phi'^{}_{e}\rangle\phi'^{}_{e}+\langle\phi'^{}_{\nu}\rangle\phi'^{}_{\nu}}
{\sqrt{\langle\phi'^{}_{d}\rangle^2_{}+\langle\phi'^{}_{u}\rangle^2_{}+\langle\phi'^{}_{e}\rangle^2_{}+\langle\phi'^{}_{\nu}\rangle^2_{}}}\,,
\end{eqnarray}
which will be responsible for spontaneously breaking the left-right symmetry down to the electroweak symmetry, i.e
\begin{eqnarray}
\label{lrsb}
SU(2)_L^{}\times SU(2)^{}_{R}\times U(1)^{}_{B-L}\stackrel{\langle\varphi'\rangle}{\longrightarrow} SU(2)_L^{}\times U(1)^{}_{Y}\,.
\end{eqnarray}
As for the $[SU(2)^{}_L]$-doublet Higgs scalars $\phi_{d,u,e,\nu}^{}$, their quadratic terms are
\begin{eqnarray}
V\supset \sum_{a=d,u,e,\nu}^{}\mu^2_a \phi^\dagger_a\phi^{}_a + \sum_{a\neq b}^{}\mu^2_{ab} \phi^\dagger_a \phi^{}_b~~(\mu^2_{ab}=\mu^{2}_{ba})\,,
\end{eqnarray}
and hence they can form
\begin{eqnarray}
\varphi=\frac{\langle\phi^{}_{d}\rangle \phi^{}_{d}+\langle\phi^{}_{u}\rangle\phi^{}_{u}+\langle\phi^{}_{e}\rangle\phi^{}_{e}+\langle\phi^{}_{\nu}\rangle\phi^{}_{\nu}}
{\sqrt{\langle\phi^{}_{d}\rangle^2_{}+\langle\phi^{}_{u}\rangle^2_{}+\langle\phi^{}_{e}\rangle^2_{}+\langle\phi^{}_{\nu}\rangle^2_{}}}\,,
\end{eqnarray}
to drive the spontaneous electroweak symmetry breaking,
\begin{eqnarray}
\label{ewsb}
SU(2)_L^{}\times U(1)^{}_{Y}\stackrel{\langle\varphi\rangle}{\longrightarrow} U(1)^{}_{em}\,.
\end{eqnarray}
We also assume the discrete parity symmetry has been spontaneously or softly broken before the gauge symmetry breaking (\ref{lrsb}) and (\ref{ewsb}). So, the vacuum expectation values (VEVs) can be chosen by
\begin{eqnarray}
\langle\phi_{\nu}^{}\rangle\ll\langle\phi_{d,u,e}^{}\rangle\ll \langle\phi'^{}_{\nu}\rangle\sim\langle\phi'^{}_{d,u,e}\rangle\,.
\end{eqnarray}
Here the smallness of the VEV $\langle\phi_{\nu}^{}\rangle$ may be understood by a seesaw suppression,
 \begin{eqnarray}
\langle\phi_{\nu}^{}\rangle&\simeq& -\frac{\sum_{a\neq \nu}^{}\mu^2_{\nu a}\langle\phi_{a}^{}\rangle}{\mu_\nu^2}\ll \langle\phi_{d,u,e}^{}\rangle\nonumber\\
&&\textrm{for}~~\mu_\nu^2\gg \mu^2_{\nu d,\nu u,\nu e}\,.
\end{eqnarray}
On the other hand, the $Z^{}_3$ and $Z'^{}_3$ symmetry will be conserved at any scales. This means the two scalars $\xi$ and $\chi$ which have a cubic term,
\begin{eqnarray}
\mathcal{L}\supset-\rho(\xi^\ast_{}\chi^2_{}+\textrm{H.c.})\,,
\end{eqnarray}
will not acquire any nonzero VEVs.

The full Yukawa interactions are parity invariant  
\begin{eqnarray}
\label{yukawa}
\!\!\!\!\mathcal{L}_Y^{}\!\!&=&\!\! -y_{d}^{}(\bar{q}^{}_L\tilde{\phi}^{}_dd^{}_R+\bar{q}'^{}_R\tilde{\phi}'^{}_d d'^{}_L)
-y_{u}^{}(\bar{q}^{}_L\phi^{}_u u^{}_R+\bar{q}'^{}_R\phi'^{}_u u'^{}_L)\nonumber\\
[2mm]
\!\!\!\!\!\!&&\!\!-y_{e}^{}(\bar{l}^{}_L\tilde{\phi}^{}_e e^{}_R+\bar{l}'^{}_R\tilde{\phi}'^{}_e e'^{}_L)
-y_{\nu}^{}(\bar{l}^{}_L\phi^{}_\nu \nu^{}_R+\bar{l}'^{}_R\phi'^{}_\nu \nu'^{}_L)\nonumber\\
[2mm]
\!\!\!\!\!\!&&\!\!-f^{}_{d}\chi\bar{d}^{}_Rd'^{}_L-f^{}_{u}\chi\bar{u}^{}_Ru'^{}_L
-f^{}_{e}\xi\bar{e}^{}_Re'^{}_L+\textrm{H.c.}\nonumber\\
[2mm]
\!\!\!\!\!\!&&\!\!\textrm{with}~~f^{}_{d,u,e}=f^\dagger_{d,u,e}\,,
\end{eqnarray}
which exactly respect the $[Z^{}_4\times Z^{}_3]\times [Z'^{}_4\times Z'^{}_3]$ discrete symmetries. Note the gauge-invariant Majorana and Dirac mass terms involving the ordinary right-handed neutrinos $\nu^{}_R$ and the mirror left-handed neutrinos $\nu'^{}_L$ will not appear due to the $Z^{}_3\times Z'^{}_3$ symmetries. So, it is easy to read the following relation between the ordinary and mirror fermion mass matrices,
\begin{eqnarray}
\label{mofmass}
\frac{\langle\phi'^{}_d\rangle}{\langle\phi^{}_d\rangle}&=&\frac{M_{d'}^{}}{m_{d}^{}}
=\frac{M_{s'}^{}}{m_{s}^{}}=\frac{M_{b'}^{}}{m_{b}^{}}\,,\nonumber\\
[2mm]
\frac{\langle\phi'^{}_u\rangle}{\langle\phi^{}_u\rangle}&=&\frac{M_{u'}^{}}{m_{u}^{}}
=\frac{M_{c'}^{}}{m_{c}^{}}=\frac{M_{t'}^{}}{m_{t}^{}}~~\textrm{with}\nonumber\\
[2mm]
&&V=V'^{}_{\textrm{CKM}}=V^{}_{\textrm{CKM}}\,;\nonumber\\
[4mm]
\frac{\langle\phi'^{}_e\rangle}{\langle\phi^{}_e\rangle}&=&\frac{M_{e'}^{}}{m_{e}^{}}=\frac{M_{\mu'}^{}}{m_{\mu}^{}}
=\frac{M_{\tau'}^{}}{m_{\tau}^{}}\,,\nonumber\\
[2mm]
\frac{\langle\phi'^{}_\nu\rangle}{\langle\phi^{}_\nu\rangle}
&=&\frac{M_{\nu'^{}_{1}}^{}}{m_{1}^{}}=\frac{M_{\nu'^{}_{2}}^{}}{m_{2}^{}}
=\frac{M_{\nu'^{}_{3}}^{}}{m_{3}^{}}~~\textrm{with}\nonumber\\
[2mm]
&&U=U'^{}_{\textrm{PMNS}}=U^{}_{\textrm{PMNS}}\,.
\end{eqnarray}
Here $m_{f}^{}$ and $M_{f'}^{}$ stand for the ordinary and mirror fermion mass eigenvalues, while $V$ and $U$ are the CKM and PMNS matrices \cite{patrignani2016}. Clearly, the mirror Dirac neutrinos have a heavy mass matrix proportional to the light mass matrix of the ordinary Dirac neutrinos. The PMNS matrix also does not contain any Majorana CP phases,
\begin{widetext}
\begin{eqnarray}\label{pmns}
U=\left[\begin{array}{ccccl}
c_{12}^{}c_{13}^{}&& s_{12}^{}c_{13}^{}&&  s_{13}^{}e^{-i\delta}_{}\\
[4mm] -s_{12}^{}c_{23}^{}-c_{12}^{}s_{23}^{}s_{13}^{}e^{i\delta}_{}
&&~~c_{12}^{}c_{23}^{}-s_{12}^{}s_{23}^{}s_{13}^{}e^{i\delta}_{}
&& s_{23}^{}c_{13}^{}\\
[4mm] ~~s_{12}^{}s_{23}^{}-c_{12}^{}c_{23}^{}s_{13}^{}e^{i\delta}_{}
&& -c_{12}^{}s_{23}^{}-s_{12}^{}c_{23}^{}s_{13}^{}e^{i\delta}_{}
&& c_{23}^{}c_{13}^{}
\end{array}\right].
\end{eqnarray}
\end{widetext}

For the following demonstration, we denote the lightest mass eigenvalue of the mirror(ordinary) neutrino mass matrix by
\begin{eqnarray}
\label{spectrum}
M_{\nu'^{}_x}^{}&=&\textrm{min}\{M_{\nu'^{}_1}^{}\,,~M_{\nu'^{}_2}^{}\,,~M_{\nu'^{}_3}^{}\}\,,\nonumber\\
[2mm]
m_{\nu^{}_x}^{}&=&\textrm{min}\{m_{1}^{}\,,~m_{2}^{}\,,~m_{3}^{}\}\,.
\end{eqnarray}
Note we have the flexibility to choose the mirror VEVs $\langle\phi'^{}_{d,u,e,\nu}\rangle$. Therefore, at least one generation of mirror neutrinos can be in the range as below,
\begin{eqnarray}
\label{spectrum}
M_{\nu'^{}_i}^{}&\gg& M_{\mu'}^{}+M_{d'}^{}+M_{u'}^{}\,,\nonumber\\
[2mm]
M_{\nu'^{}_i}^{}&<& M_{\tau'}^{}+M_{d'}^{}+M_{u'}^{}\,, ~M_{W_R^{}}^{}+M_{e'}^{}\,,
\end{eqnarray}
where the $W_R^{\pm}$ mass is given by
\begin{eqnarray}
\label{mowmass}
M_{W^{}_R}^{}=\frac{g}{\sqrt{2}}\langle\varphi'\rangle\,.
\end{eqnarray}

\section{Mirror lepton asymmetries}

\begin{figure*}
\vspace{7cm} \epsfig{file=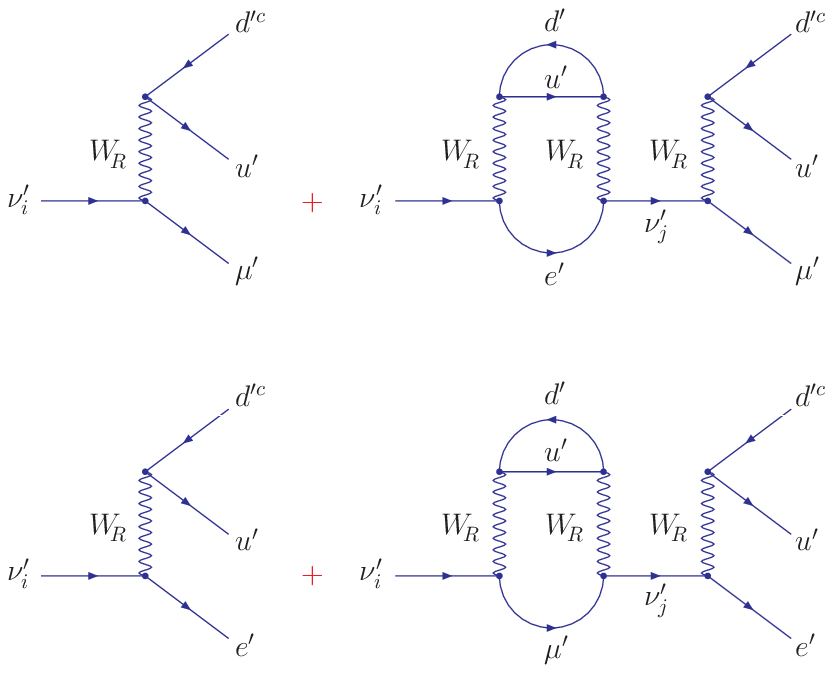, bbllx=2.25cm, bblly=6.0cm,
bburx=12.25cm, bbury=16cm, width=7.5cm, height=7.5cm, angle=0,
clip=0} \vspace{-6cm} \caption{\label{mndecay} The lepton-number-conserving decays for generating a lepton asymmetry in the mirror muons and an opposite lepton asymmetry in the mirror electrons.  }
\end{figure*}

For a mirror neutrino in the range (\ref{spectrum}), it can have the three-body decay modes as shown in Fig. \ref{mndecay}. We calculate the decay width at tree level,
\begin{eqnarray}
\label{width1}
\Gamma_{\nu'^{}_i}^{}&=&\Gamma(\nu'^{}_i\rightarrow
e'+u'+d'^c_{})+\Gamma(\nu'^{}_i\rightarrow
\mu'+u'+d'^c_{})
\nonumber\\
[2mm]
&\simeq&\frac{g^4_{}}{2^{11}_{}\pi^3_{}}\frac{M_{\nu'^{}_{i}}^{5}}{M_{W_R^{}}^4}\left(|U_{ei}^{}|^2_{}+|U_{\mu
i}^{}|^2_{}\right)~~\textrm{for}~~M_{\nu'^{}_i}^2\ll M_{W_R^{}}^2\,.\nonumber\\
&&\end{eqnarray}
For convenience, we define the parameter 
\begin{eqnarray}
r\equiv \frac{M_{\nu'^{}_x}^{2}}{M_{W_R^{}}^{2}}\,,
\end{eqnarray}
and then rewrite the decay width (\ref{width1}) by 
\begin{eqnarray}
\label{width2}
\Gamma_{\nu'^{}_i}^{}&\simeq&\frac{g^4_{}}{2^{11}_{}\pi^3_{}}\left(|U_{ei}^{}|^2_{}+|U_{\mu
i}^{}|^2_{}\right)\frac{M_{\nu_i^{}}^5}{M_{\nu'^{}_x}^5}\frac{M_{\nu'^{}_{x}}^{5}}{M_{W_R^{}}^{4}}\nonumber\\
&=&\frac{g^4_{}}{2^{11}_{}\pi^3_{}}\left(|U_{ei}^{}|^2_{}+|U_{\mu
i}^{}|^2_{}\right)\frac{m_i^5}{m_{\nu_x^{}}^5}M_{\nu'^{}_{x}}^{}r^2_{}\,.
\end{eqnarray}
Although the above mirror neutrino decays exactly conserve the lepton number, they can generate a lepton asymmetry in the mirror muons and an opposite lepton asymmetry in the mirror electrons,
\begin{subequations}
\begin{eqnarray}
\varepsilon_{\nu'^{}_{i}}^{\mu'}\!&=&\!\frac{\Gamma(\nu'^{}_i\rightarrow
\mu'^{-}_{}+u'+d'^c_{})-\Gamma(\nu'^{c}_{i}\rightarrow
\mu'^{+}_{}+u'^c_{}+d')}{\Gamma_{\nu'^{}_i}^{}}\,,\nonumber\\
&&\\
[2mm]
\varepsilon_{\nu'^{}_{i}}^{e'}\!&=&\!\frac{\Gamma(\nu'^{}_i\rightarrow
e'^{-}_{}+u'+d'^c_{})-\Gamma(\nu'^{c}_i\rightarrow
e'^{+}_{}+u'^c_{}+d')}{\Gamma_{\nu'^{}_i}^{}}\,.\nonumber\\
&&
\end{eqnarray}
\end{subequations}
The CP asymmetries $\varepsilon_{\nu'^{}_{i}}^{\mu'}$ and $\varepsilon_{\nu'^{}_{i}}^{e'}$ can be evulated at one-loop level,
\begin{eqnarray}
\varepsilon_{\nu'^{}_{i}}^{\mu'}&=&-\varepsilon_{\nu'^{}_{i}}^{e'}\nonumber\\
&\simeq&\frac{g^4_{}}{2^9_{}\pi^3_{}}\sum_{j\neq i}^{}\frac{\textrm{Im}( U_{\mu i}^{}U_{e i}^{\ast} U_{e j}^{} U_{\mu j}^{\ast} )}{|U_{\mu i}^{}|^2_{}+|U_{e i}^{}|^2_{}}
\frac{M_{\nu'^{}_i}^2}{M_{\nu'^{}_j}^2-M_{\nu'^{}_i}^2}\frac{M_{\nu'^{}_i}^4}{M_{W_R^{}}^4}\nonumber\\
&=&\frac{g^4_{}}{2^{9}_{}\pi^3_{}}\sum_{j\neq i}^{}\frac{\textrm{Im}(U_{\mu i}^{}U_{e i}^{\ast} U_{e j}^{} U_{\mu j}^{\ast})}{|U_{\mu i}^{}|^2_{}+|U_{e i}^{}|^2_{}}
\frac{m_i^2}{\Delta m_{ji}^2}\frac{M_{\nu'^{}_i}^4}{M_{\nu'^{}_x}^4}\frac{M_{\nu'^{}_x}^4}{M_{W_R^{}}^4}\nonumber\\
&=&\frac{g^4_{}}{2^{9}_{}\pi^3_{}}\sum_{j\neq i}^{}\frac{\textrm{Im}(U_{\mu i}^{}U_{e i}^{\ast} U_{e j}^{} U_{\mu j}^{\ast})}{|U_{\mu i}^{}|^2_{}+|U_{e i}^{}|^2_{}}
\frac{m_i^2}{\Delta m_{ji}^2}\frac{m_i^4}{m_{\nu_x^{}}^4}r^2_{}\,,
\end{eqnarray}
with $\Delta m_{ji}^2\equiv m_j^2-m_i^2$ being the squared mass differences of the ordinary neutrinos. Remarkably, the above CP asymmetries are fully determined by the neutrino mass eigenvalues and the PMNS matrix, up to an overall factor $r^2_{}$. By inserting the PMNS matrix (\ref{pmns}), we then can read
\begin{widetext}
\begin{eqnarray}
\label{cpas}
\varepsilon_{\nu'^{}_{1}}^{\mu'}=-\varepsilon_{\nu'^{}_{1}}^{e'}
&= &\frac{g^4_{}}{2^{9}_{}\pi^3_{}}\frac{J_{CP}^{}}
{c_{12}^2c_{13}^2+s_{12}^2c_{23}^2+c_{12}^2s_{23}^2s_{13}^2+2s_{12}^{}c_{12}^{}s_{23}^{}c_{23}^{}s_{13}^{}\cos\delta} \left(\frac{m_1^2}{\Delta m_{21}^2}+\frac{m_1^2}{\Delta m_{31}^2}\right)\frac{m_{\nu^{}_1}^{4}}{m_{\nu_x^{}}^{4}}r^2_{}\,,\nonumber\\
[2mm]
\varepsilon_{\nu'^{}_{2}}^{\mu'}=-\varepsilon_{\nu'^{}_{2}}^{e'}
&= & \frac{g^4_{}}{2^{9}_{}\pi^3_{}}\frac{J_{CP}^{}}
{s_{12}^2c_{13}^2+c_{12}^2c_{23}^2+s_{12}^2s_{23}^2s_{13}^2-2s_{12}^{}c_{12}^{}s_{23}^{}c_{23}^{}s_{13}^{}\cos\delta}\left(\frac{m_2^2}{\Delta m_{21}^2}-\frac{m_2^2}{\Delta m_{32}^2}\right)\frac{m_{\nu^{}_2}^{4}}{m_{\nu_x^{}}^{4}}r^2_{}\,,\nonumber\\
[2mm]
\varepsilon_{\nu'^{}_{3}}^{\mu'}=-\varepsilon_{\nu'^{}_{3}}^{e'}
&= &\frac{g^4_{}}{2^{9}_{}\pi^3_{}}\frac{J_{CP}^{}}{s_{13}^2+s_{23}^2c_{13}^2}\left(\frac{m_3^2}{\Delta m_{31}^2}-\frac{m_3^2}{\Delta m_{32}^2}\right) \frac{m_{\nu^{}_3}^{4}}{m_{\nu_x^{}}^{4}}r^2_{}\,.
\end{eqnarray}
\end{widetext}
Here we have quoted the CP-violating parameter \cite{jarlskog1985},
\begin{eqnarray}
J_{CP}^{}=s_{12}^{}c_{12}^{}s_{23}^{}c_{23}^{}s_{13}^{} c_{13}^{2}\sin\delta\,.
\end{eqnarray}

\begin{figure*}
\vspace{10cm} \epsfig{file=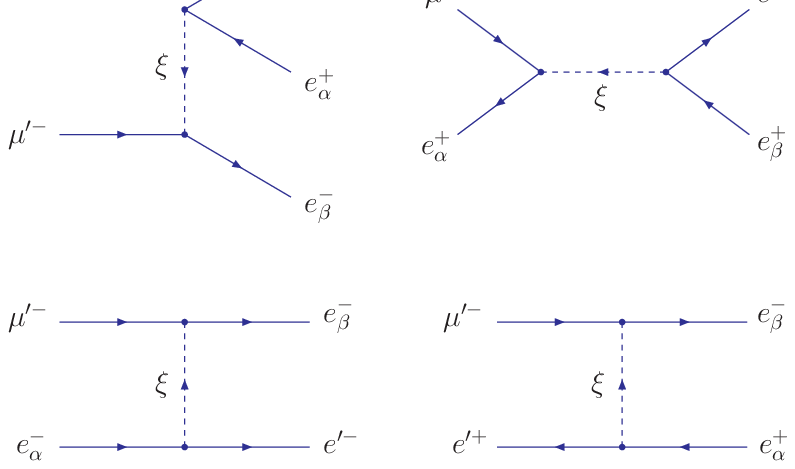, bbllx=5.8cm, bblly=6.0cm,
bburx=15.8cm, bbury=16cm, width=7.5cm, height=7.5cm, angle=0,
clip=0} \vspace{-10.5cm} \caption{\label{conversation} The conversations between the mirror muons and the mirror electrons. The CP-conjugation processes are not shown for simplicity.}
\end{figure*}

In the presence of the scalar $\xi$, the induced mirror electron and muon asymmetries will be immediately cancelled each other if the decaying and scattering processes shown in Fig. \ref{conversation} are very fast. We expect such processes to go into equilibrium at some low temperatures such as $T<T_{\textrm{sph}}^{}\sim 100\,\textrm{GeV}$ where the $SU(2)^{}_L$ sphaleron processes have not been active no longer. For this purpose we can require the interaction rates smaller than the Hubble constant at the crucial temperature $T_{\textrm{sph}}^{}$,
\begin{eqnarray}
\label{oe-conversation1}
\!\!\!\!\!\!\!\!\Gamma_{D}^{\mu'\rightarrow e'}\!&=&\!\sum_{\alpha\beta}^{}\Gamma(\mu'^{-}_{}\rightarrow
e'^{-}_{}+e^{-}_{\alpha}+e^{+}_{\beta})\nonumber\\
\!\!\!\!\!\!\!\!\!&=&\!\frac{\left(f^{2}_e\right)_{\mu'\mu'}^{}
\left(f^{2}_e\right)_{e'e'}^{}}{3\cdot 2^{11}_{}\pi^3_{}}\frac{M_{\mu'}^{5}}{M_{\xi}^4}\nonumber\\
\!\!\!\!\!\!\!\!\!&=&\!\frac{\left(f^{2}_e\right)_{\mu'\mu'}^{}
\left(f^{2}_e\right)_{e'e'}^{}}{3\cdot 2^{11}_{}\pi^3_{}}\frac{m_\mu^5}{m_e^5}\frac{M_{e'}^{5}}{M_{\xi}^4}\nonumber\\
\!\!\!\!\!\!\!\!\!&\simeq&\!2\times 10^{-21}_{}\,\textrm{GeV}\left(\frac{M_{e'}^{}}{10^7_{}\,\textrm{GeV}}\right)^5_{}\left(\frac{10^{12}_{}\,\textrm{GeV}}{M_{\xi}^{}}\right)^4_{}\nonumber\\
\!\!\!\!\!\!\!\!\!&&\!\times\left[\frac{\left(f^{2}_e\right)_{\mu'\mu'}^{}}{10^{-7}_{}}\right]
\left[\frac{\left(f^{2}_e\right)_{e'e'}^{}}{10^{-7}_{}}\right]
<H(T)\left|^{}_{T_{\textrm{sph}}}\right.,
\end{eqnarray}
\begin{eqnarray}
\label{oe-conversation2}
\!\!\!\!\!\!\!\!\Gamma_{s}^{\mu'\rightarrow e'}\!&=&\!
\sum_{\alpha\beta}^{}\Gamma(\mu'^{-}_{}+e^{+}_{\alpha}\rightarrow
e'^{-}_{}+e^{+}_{\beta})\nonumber\\
\!\!\!\!\!\!\!\!\!&=&\!\frac{3\left(f^{2}_e\right)_{\mu'\mu'}^{}
\left(f^{2}_e\right)_{e'e'}^{}}{4\pi^3_{}}\frac{T_{}^{5}}{M_{\xi}^4}\nonumber\\
\!\!\!\!\!\!\!\!\!&=&\!2.4\times 10^{-54}_{}\,\textrm{GeV}\left(\frac{T}{100\,\textrm{GeV}}\right)^5_{}\left(\frac{10^{12}_{}\,\textrm{GeV}}{M_{\xi}^{}}\right)^4_{}\nonumber\\
\!\!\!\!\!\!\!\!\!&&\!\times\left[\frac{\left(f^{2}_e\right)_{\mu'\mu'}^{}}{10^{-7}_{}}\right]
\left[\frac{\left(f^{2}_e\right)_{e'e'}^{}}{10^{-7}_{}}\right]
<H(T)\left|^{}_{T_{\textrm{sph}}}\right.,
\end{eqnarray}
\begin{eqnarray}
\label{oe-conversation3}
\!\!\!\!\!\!\!\!\Gamma_{t}^{\mu'\rightarrow e'}\!&=&\!\sum_{\alpha\beta}^{}\Gamma(\mu'^{-}_{}+e^{-}_{\alpha}\rightarrow
e'^{-}_{}+e^{-}_{\beta})\nonumber\\
\!\!\!\!\!\!\!\!\!&=&\!\frac{\left(f^\dagger_{e}f^{}_e\right)_{\mu'\mu'}^{}
\left(f^{2}_e\right)_{e'e'}^{}}{4\pi^3_{}}\frac{T_{}^{5}}{M_{\xi}^4}\nonumber\\
\!\!\!\!\!\!\!\!\!&=&\!8.1\times 10^{-55}_{}\,\textrm{GeV}\left(\frac{T}{100\,\textrm{GeV}}\right)^5_{}\left(\frac{10^{12}_{}\,\textrm{GeV}}{M_{\xi}^{}}\right)^4_{}\nonumber\\
\!\!\!\!\!\!\!\!\!&&\!\times
\left[\frac{\left(f^{2}_e\right)_{\mu'\mu'}^{}}{10^{-7}_{}}\right]
\left[\frac{\left(f^{2}_e\right)_{e'e'}^{}}{10^{-7}_{}}\right]
<H(T)\left|^{}_{T_{\textrm{sph}}}\right.,
\end{eqnarray}
\begin{eqnarray}
\label{oe-conversation4}
\!\!\!\!\!\!\!\!\Gamma_{t}^{\mu'+e'}\!&=&\!\sum_{\alpha\beta}^{}\Gamma(\mu'^{-}_{}+e^{+}_{\alpha}\rightarrow
e'^{-}_{}+e^{+}_{\beta})\nonumber\\
\!\!\!\!\!\!\!\!\!&=&\!\frac{\left(f^{2}_e\right)_{\mu'\mu'}^{}
\left(f^{2}_e\right)_{e'e'}^{}}{4\pi^3_{}}\frac{T_{}^{5}}{M_{\xi}^4}\nonumber\\
\!\!\!\!\!\!\!\!\!&=&\!8.1\times 10^{-55}_{}\,\textrm{GeV}\left(\frac{T}{100\,\textrm{GeV}}\right)^5_{}\left(\frac{10^{12}_{}\,\textrm{GeV}}{M_{\xi}^{}}\right)^4_{}\nonumber\\
\!\!\!\!\!\!\!\!\!&&\!\times
\left[\frac{\left(f^{2}_e\right)_{\mu'\mu'}^{}}{10^{-7}_{}}\right]
\left[\frac{\left(f^{2}_e\right)_{e'e'}^{}}{10^{-7}_{}}\right]
<H(T)\left|^{}_{T_{\textrm{sph}}}\right..
\end{eqnarray}
Here the Hubble constant is given by
\begin{eqnarray}
H(T)&=&\left(\frac{8\pi^{3}_{}g_{\ast}^{}}{90}\right)^{\frac{1}{2}}_{}
\frac{T^{2}_{}}{M_{\textrm{Pl}}^{}}\nonumber\\
&=&1.4\times 10^{-14}_{}\,\textrm{GeV}\left(\frac{T}{100\,\textrm{GeV}}\right)^2_{}\,,
\end{eqnarray}
with $M_{\textrm{Pl}}^{}\simeq 1.22\times 10^{19}_{}\,\textrm{GeV}$ being the Planck mass and $g_{\ast}^{}=\mathcal{O}(100)$ being the relativistic degrees of freedom.

Furthermore, like the ordinary $\mu\rightarrow e\gamma$ process, the mirror muon will decay into the mirror electron and the photon at one-loop level,
\begin{eqnarray}
\Gamma(\mu'\rightarrow e'\gamma)&=&\frac{\alpha g^4_{}}{2^{16}_{}\pi^{4}_{}}\frac{M_{\mu'}^{5}}{M_{W_R^{}}^{4}}|\delta_\nu^{}|^2_{}\nonumber\\
&=&\frac{\alpha g^4_{}}{2^{16}_{}\pi^{4}_{}}\frac{M_{\mu'}^5}{M_{\tau'}^4 M_{e'}^{}}\frac{M_{\tau'}^4 M_{e'}^{}}{M_{\nu'^{}_x}^{4}}\frac{M_{\nu'^{}_x}^{4}}{M_{W_R^{}}^{4}}|\delta_\nu^{}|^2_{}\nonumber\\
&=&\frac{\alpha g^4_{}}{2^{16}_{}\pi^{4}_{}}\frac{m_\mu^5}{m_\tau^4 m_e^{}}|\delta_\nu^{}|^2_{}r^2_{} \frac{M_{\tau'}^4 M_{e'}^{}}{M_{\nu'^{}_x}^4}\nonumber\\
&\leq&\frac{\alpha g^4_{}}{2^{16}_{}\pi^{4}_{}}\frac{m_\mu^5}{m_\tau^4 m_e^{}}|\delta_\nu^{}|^2_{}r^2_{} M_{e'}^{}\,.
\end{eqnarray}
Here $\delta_\nu^{}$ is the GIM suppression factor,
\begin{eqnarray}
\delta_\nu^{}&=&\frac{\sum_{i}^{}U_{ei}^{} U_{\mu i}^{\ast}M_{\nu'^{}_i}^2}{M_{W_R^{}}^2}\nonumber\\
&=&\frac{M_{\nu'^{}_x}^2}{M_{W_R^{}}^2}\frac{\sum_{i}^{}U_{ei}^{} U_{\mu i}^{\ast}M_{\nu'^{}_i}^2}{M_{\nu'^{}_x}^2}\nonumber\\
&=&\frac{r}{m_{\nu_x^{}}^2}[s_{12}^{}c_{12}^{}c_{23}^{}c_{13}^{}\Delta m_{21}^2\nonumber\\
&&+s_{23}^{}s_{13}^{}c_{13}^{}(\Delta m_{32}^2 s_{12}^2 +\Delta m_{31}^2 c_{12}^2)e^{-i\delta}_{}]\,.
\end{eqnarray}
We hence can require
\begin{eqnarray}
\label{mmeg}
M_{e'}^{}<4\times 10^{10}_{}\,\textrm{GeV}\left(\frac{m_{\nu_x^{}}^{}}{0.2\,\textrm{eV}}\right)^4_{}\left(\frac{0.1}{r}\right)^{4}_{}\,,
\end{eqnarray}
to guarantee
\begin{eqnarray}
\Gamma(\mu'\rightarrow e'\gamma)<H(T)\left|^{}_{T_{\textrm{sph}}}\right.\,.
\end{eqnarray}
In the above estimations, we have input the charged lepton masses $m_e^{}=511\,\textrm{keV}$, $m_\mu^{}=106\,\textrm{MeV}$, $m_\tau^{}=1.78\,\textrm{GeV}$, 
as well as the neutrino oscillation data $\Delta m_{31}^2 =2.45\times 10^{-3}_{}\,\textrm{eV}^2_{}$, $\Delta m_{21}^2=7.53\times 10^{-5}_{}\,\textrm{eV}^2_{}$, $\sin^2_{}\theta_{12}^{}=0.304$, $\sin^2_{}\theta_{23}^{}=0.51$, $\sin^2_{}\theta_{13}^{}=0.0219$ \cite{patrignani2016}.

\section{Ordinary baryon asymmetry}

\begin{figure*}
\vspace{7.5cm} \epsfig{file=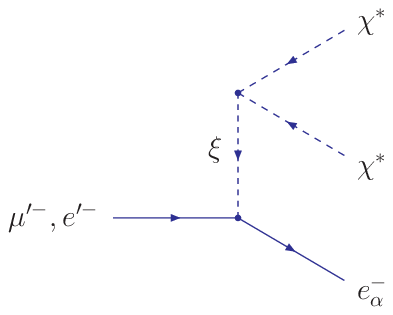, bbllx=5cm, bblly=6.0cm,
bburx=15cm, bbury=16cm, width=7.5cm, height=7.5cm, angle=0,
clip=0} \vspace{-10.5cm} \caption{\label{emudecay} The three-body decays of the mirror muons and electrons into the ordinary leptons and the dark matter scalar.}
\end{figure*}

It is well known we can obtain a baryon asymmetry from a lepton asymmetry produced before the $SU(2)^{}_L$ sphaleron processes stop working at a temperature around $T_{\textrm{sph}}^{}\sim 100\,\textrm{GeV}$ \cite{krs1985}. Now we have already got a lepton asymmetry stored in the mirror muons and an opposite lepton asymmetry stored in the mirror electrons. These mirror muons and electrons will decay into the ordinary right-handed leptons with some stable scalars. The mirror muon and electron decays have been shown in Fig. \ref{emudecay}. If the mirror muons and electrons both decay efficiently above the scale $T_{\textrm{sph}}^{}$, the mirror muon and electron asymmetries will both participate in the sphalerons. In consequence, we will fail in getting a nonzero baryon asymmetry from the induced mirror lepton asymmetries. However, it is allowed that the mirror muons can have a shorter life time while the mirror electrons can have a longer life time, i.e.
\begin{eqnarray}
\label{mmlife}
\Gamma_{\mu'}^{}\!&=&\!\sum_{\alpha}^{}\left[\Gamma(\mu'^{-}_{}\rightarrow
e^{-}_{\alpha}+\chi^\ast_{}+\chi^\ast_{})\right]=\frac{\left(f^{2}_e\right)_{\mu'\mu'}^{}}{3\cdot 2^{9}_{}\pi^3_{}}
\frac{M_{\mu'}^{3}\rho^2_{}}{M_{\xi}^4}\nonumber\\
\!&=&\!
1.9\times 10^{-8}_{}\,\textrm{GeV}\left(\frac{M_{e'}^{}}{10^7_{}\,\textrm{GeV}}\right)^3_{}\left(\frac{10^{12}_{}\,\textrm{GeV}}{M_{\xi}^{}}\right)^4_{}\nonumber\\
\!&&\!\times\left(\frac{\rho}{M_\xi^{}}\right)^2_{}
\left[\frac{\left(f^{2}_e\right)_{\mu'\mu'}^{}}{10^{-7}_{}}\right] >H(T_{\textrm{sph}}^{})\,,\\
[2mm]
\label{melife}
\Gamma_{e'}^{}\!&=&\!\sum_{\alpha}^{}\left[\Gamma(e'^{-}_{}\rightarrow
e^{-}_{\alpha}+\chi^\ast_{}+\chi^\ast_{})\right]=\frac{\left(f^{2}_e\right)_{e'e'}^{}}{3\cdot 2^{9}_{}\pi^3_{}}\frac{M_{e'}^{3}\rho^2_{}}{M_{\xi}^4}\nonumber\\
\!&=&\!
2.1\times 10^{-15}_{}\,\textrm{GeV}\left(\frac{M_{e'}^{}}{10^{7}_{}\,\textrm{GeV}}\right)^3_{}\left(\frac{10^{12}_{}\,\textrm{GeV}}{M_{\xi}^{}}\right)^4_{}
\nonumber\\
\!&&\!\times
\left(\frac{\rho}{M_\xi^{}}\right)^2_{}\left[\frac{\left(f^{2}_e\right)_{e'e'}^{}}{10^{-7}_{}}\right] <H(T_{\textrm{sph}}^{})\,.
\end{eqnarray}
In this case, similar to the left-handed lepton asymmetry and the opposite right-handed neutrino asymmetry in the neutrinogenesis scenario \cite{dlrw1999}, the mirror muon asymmetry rather than the mirror electron asymmetry will be partially converted to an ordinary baryon asymmetry through the sphalerons.

In the weak washout region where
\begin{eqnarray}
K_i^{}&=&\frac{\Gamma_{\nu'^{}_i}^{}}{2H(T)}\left|_{T=M_{\nu'^{}_i}^{}}^{}\right.\ll 1\,,
\end{eqnarray}
we have \cite{kt1990},
\begin{eqnarray}
\label{ba}
\!\!\!\!\!\!\!\!\!\!\!\!&&\frac{\eta_B^{}}{7.04}\!\simeq\!-\frac{28}{79}\!\times \! \left\{\begin{array}{cl}\frac{\varepsilon_{\nu'^{}_1}^\mu+\varepsilon_{\nu'^{}_2}^\mu
+\varepsilon_{\nu'^{}_3}^\mu}{g_\ast^{}}&\textrm{for}~~M_{\nu'^{}_{1}}^{}\simeq M_{\nu'^{}_{2}}^{}\simeq M_{\nu'^{}_{3}}^{}\,,\\
[3mm]
\frac{\varepsilon_{\nu'^{}_1}^\mu+\varepsilon_{\nu'^{}_2}^\mu}{g_\ast^{}} &\textrm{for} ~~M_{\nu'^{}_{1}}^{}\simeq M_{\nu'^{}_{2}}^{}\ll M_{\nu'^{}_{3}}^{}\,,\\
[3mm]
\frac{\varepsilon_{\nu'^{}_1}^\mu}{g_\ast^{}} &\textrm{for} ~~M_{\nu'^{}_{1}}^{}\ll M_{\nu'^{}_{2}}^{}< M_{\nu'^{}_{3}}^{}\,,\\
[3mm]
\frac{\varepsilon_{\nu'^{}_3}^\mu+\varepsilon_{\nu'^{}_1}^\mu}{g_\ast^{}}&\textrm{for}~~M_{\nu'^{}_{3}}^{}\simeq M_{\nu'^{}_{1}}^{}\ll M_{\nu'^{}_{2}}^{}\\
[3mm]
\frac{\varepsilon_{\nu'^{}_3}^\mu}{g_\ast^{}}&\textrm{for}~~M_{\nu'^{}_{3}}^{}\ll M_{\nu'^{}_{1}}^{}< M_{\nu'^{}_{2}}^{}\,.\end{array}\right.\nonumber\\
\!\!\!\!\!\!\!\!\!\!\!\!&&
\end{eqnarray}
The weak washout condition can be achieved by
\begin{eqnarray}
\label{conww}
\!\!K_i^{}&=& 0.1\left(\frac{r}{0.1}\right)^2_{}\left(\frac{10^{11}_{}\,\textrm{GeV}}{M_{\nu'^{}_x}^{}}\right) (|U_{ei}|^2_{}+|U_{\mu i}^{}|^2_{})\frac{m_i^5}{m_{\nu_x^{}}^5}\nonumber\\
\!\!&\ll& 1\,.
\end{eqnarray}

When Eqs. (\ref{oe-conversation1}-\ref{oe-conversation4}), (\ref{mmeg}), (\ref{mmlife}-\ref{melife}) are satisfied, we only need four parameters $M_{\nu'^{}_x}^{}$, $r$, $m_{\nu_x^{}}^{}$ and $\sin\delta$ to determine the final baryon asymmetry $\eta_B^{}$ by Eqs. (\ref{cpas}), (\ref{ba}) and (\ref{conww}). If Eq. (\ref{conww}) is further satisfied, the CP asymmetry (\ref{cpas}) and then the baryon asymmetry (\ref{ba}) can be described by three parameters $r$, $m_{\nu_x^{}}^{}$ and $\sin\delta$. This means the cosmic baryon asymmetry can be parametrized by the neutrino mass matrix up to an overall parameter $r$. For example, we fix $M_\xi^{}=\rho=\mathcal{O}(10^{12}_{}\,\textrm{GeV})$, $M_{e'}^{}=\mathcal{O}(10^{7}_{}\,\textrm{GeV})$ and $f^2_e=\mathcal{O}(10^{-7}_{})$ to satisfy Eqs. (\ref{oe-conversation1}-\ref{oe-conversation4}), (\ref{mmeg}), (\ref{mmlife}-\ref{melife}). We further take $M_{\nu'^{}_x}^{}=10^{11}_{}\,\textrm{GeV}$, $r=0.1$, $m_{\nu_x^{}}^{}=0.2\,\textrm{eV}$ and $\sin\delta= -0.0045$ to get $\eta_B^{}=5.91\times 10^{-10}_{}$, which is consistent to the observed value \cite{patrignani2016}.

\section{Strong CP problem and dark matter}

The present model give a non-perturbative QCD Lagrangian as follows,
\begin{eqnarray}
\!\!\!\!&&\!\!\!\!\mathcal{L}_{QCD}^{}\supset-\bar{\theta}\frac{g^2_3}{32\pi^2_{}}G\tilde{G}~~\textrm{with}~~
\bar{\theta}=\theta-\textrm{Arg}\textrm{Det} (M_u^{} M_d^{})\,,\nonumber\\
\!\!\!\!&&\!\!\!\!
\end{eqnarray}
where $\theta$ is from the QCD $\Theta$-vacuum while $M_u^{}$ and $M_d^{}$
are the mass matrices of the down-type and up-type quarks,
respectively,
\begin{eqnarray}
\mathcal{L}&\supset& -[\bar{d}^{}_L,
\bar{d}'^{}_L]M_d^{}\left[\begin{array}{c}d^{}_R\\
[2mm]
d'^{}_R\end{array}\right]-[\bar{u}^{}_L,
\bar{u}'^{}_L]M_u^{}\left[\begin{array}{c}u^{}_R\\
[2mm]
u'^{}_R\end{array}\right]+\textrm{H.c.}\nonumber\\
[2mm]
&&\textrm{with}~~M_d^{}=
\left[\begin{array}{cc}y_d^{}\langle\phi^{}_d\rangle&0\\
[2mm]
0&y_d^{\dagger}\langle\phi'^{}_d\rangle\end{array}\right],\nonumber\\
&&\quad \quad \,\, \,M_u^{}=
\left[\begin{array}{cc}y_u^{}\langle\phi^{}_u\rangle&0\\
[2mm]
0&y_u^{\dagger}\langle\phi'^{}_u\rangle\end{array}\right].
\end{eqnarray}
When the $\theta$-term is removed as a result of the
parity invariance, the real determinants
$\textrm{Det}(M_d^{})$ and $\textrm{Det}(M_u^{})$ will lead to a
zero $\textrm{Arg}\textrm{Det} (M_u^{} M_d^{})$. We hence can obtain
a vanishing strong CP phase $\bar{\theta}$ at tree level \cite{bm1989}.

The model also contains a stable scalar $\chi$ because of the unbroken $Z^{}_3\times Z'^{}_3$ symmetry. This scalar can annihilate into the ordinary species through the Higgs portal interaction. This simple dark matter scenario has been studied in a lot of literatures \cite{sz1985}.

\section{Summary}

In this paper we have proposed a new leptogenesis scenario in the left-right symmetric framework to parametrize the cosmic baryon asymmetry by the neutrino mass matrix. In our model, (i) the parity symmetry motivated to solve the strong CP problem plays an essential role, (ii) the Dirac neutrinos obtain a seesaw-suppressed mass matrix, (iii) the dark matter participates in the leptogenesis processes.

\textbf{Acknowledgement}:  This work was supported by the National Natural Science Foundation of China under Grant No. 11675100, the Recruitment Program for Young Professionals under Grant No. 15Z127060004, the Shanghai Jiao Tong University under Grant No. WF220407201, the Shanghai Laboratory for Particle Physics and Cosmology, and the Key Laboratory for Particle Physics, Astrophysics and Cosmology, Ministry of Education.

\end{document}